\documentclass[twocolumn,showpacs]{revtex4}

\usepackage{graphicx}
\usepackage{dcolumn}
\usepackage{bm}
\usepackage{hyperref}

\begin{document}

\title{Black hole evolution with the BSSN system by pseudo-spectral methods}

\author{Wolfgang Tichy}
\affiliation{Department of Physics, Florida Atlantic University,
             Boca Raton, FL  33431}




%
\newcommand\be{\begin{equation}}
\newcommand\ba{\begin{eqnarray}}

\newcommand\ee{\end{equation}}
\newcommand\ea{\end{eqnarray}}
\newcommand\p{{\partial}}
\newcommand\remove{{{\bf{THIS FIG. OR EQS. COULD BE REMOVED}}}}
%

\begin{abstract}

We present a new pseudo-spectral code for the simulation of evolution
systems that are second order in space. We test this code by evolving a
non-linear scalar wave equation. These non-linear waves can be stably
evolved using very simple constant or radiative boundary conditions,
which we show to be well-posed in the scalar wave case. The
main motivation for this work, however, is to evolve black holes for the
first time with the BSSN system by means of a spectral method. We use our
new code to simulate the evolution of a single black hole using all
applicable methods that are usually employed when the BSSN system is used
together with finite differencing methods. In particular, we use black hole
excision and test standard radiative and also constant outer boundary
conditions. Furthermore, we study different gauge choices such as $1+\log$
and constant densitized lapse. We find that these methods in principle do
work also with our spectral method. However, our simulations fail after
about $100M$ due to unstable exponentially growing modes. The reason for
this failure may be that we evolve the black hole on a full grid 
without imposing any symmetries. Such full grid instabilities have
also been observed when finite differencing methods are used to evolve
excised black holes with the BSSN system.

\end{abstract}

\maketitle

\section{Introduction}

Currently several gravitational wave detectors such as 
LIGO~\cite{LIGO_web} or GEO~\cite{GEO_web} are 
already operating, while several others are in the
planning or construction phase~\cite{Schutz99}. One of the most promising
sources for these detectors are the inspirals and mergers of binary black
holes. In order to make predictions about the final phase of such
inspirals and mergers, fully non-linear numerical simulations of the Einstein
Equations are required.
In order to numerically evolve the Einstein equations, at least
two ingredients are necessary. First we need a specific formulation
of the evolution equations. And second, a particular numerical
method is needed to implement these equations on a computer.
For both these ingredients many choices are available.
However, only a few of these choices have so far been successful
in the evolution of a binary black hole system.
In terms of commonly used evolution systems there is mainly the 
BSSN system~\cite{Baumgarte:1998te} and also the 
generalized harmonic system~\cite{Pretorius:2004jg}.
The main numerical methods are finite differencing and spectral methods.

The original generalized harmonic system
is second order and has only been used together with finite differencing
techniques~\cite{Pretorius:2004jg,Pretorius:2005gq,Pretorius:2006tp}.
Yet, recently a first order version
of this system has also been successfully evolved using a spectral
method~\cite{Lindblom:2005qh,Scheel:2006gg}.
Nevertheless, the majority of the astrophysically relevant binary black hole
simulations to date have used one particular combination:
The BSSN evolution system together with finite differencing
techniques. With this choice, several groups have recently performed
binary black hole simulations of one orbit or 
more~\cite{Bruegmann:2003aw,
Campanelli:2005dd,Campanelli:2006gf,Campanelli:2006uy,
Baker:2005vv,Baker:2006yw,Baker:2006vn}.
However, the BSSN system has never before been tested with a spectral
method. And in fact, no second order in space system such as BSSN
has ever been tried together with a spectral method. Even
for single black holes, only systems that are first order in time and 
space (such as the KST system~\cite{Kidder01a} or the 
first order version of the generalized harmonic system)
have been used successfully with a spectral method.
In order to derive such first order systems additional variables have to be
introduced which have to satisfy additional constraints. 
It is therefore possible that second order systems perform better since they
have fewer potentially unstable constraint violating solutions which can be
excited due to numerical errors.

The aim of this paper is thus to test the standard 
BSSN system with its standard boundary conditions 
in a simple case but with a pseudo-spectral method.
Such a test will show how well a spectral method will work
when employed to this widely used and successful system.
We will implement the BSSN system in its original second order
in space form, without introducing any extra variables or constraints.
This will also allow us to draw some conclusions about 
how well second order systems work when a spectral method is used.
In order to perform this task we have developed the SGRID code,
which can evolve second order in space systems using a pseudo-spectral
collocation method. This code currently supports cubical or 
spherical domains and allows for spectral expansions in Fourier
or Chebyshev basis.

Throughout we will use units where $G=c=1$.
The paper is organized as follows. 
In Sec.~\ref{codetests} we describe our methods and show results
for some simple tests where we evolve non-linear scalar waves.
Sec.~\ref{sgrid_BSSN} describes our particular implementation
of BSSN and presents tests for case of a single black hole.
We conclude with a discussion of our results in Sec.~\ref{discussion}.

\section{Description of our method and code tests}
\label{codetests}

In this section we describe the methods we use in the 
SGRID code we have developed. Furthermore, we present 
some code tests with scalar wave equations.

\subsection{The pseudo-spectral collocation method}

In one spatial dimension, spectral methods are based on expansions
\begin{equation}
u(X) = \sum_{l=1}^N \tilde{a}_l A_l(X) 
\end{equation}
of every evolved field $u(X)$ in terms of suitable basis functions $A_l(X)$
with coefficients $\tilde{a}_l$. 
Once the coefficients are known it is easy to compute
derivatives of $u(X)$ from
\begin{equation}
\partial_X u(X) = \sum_{l=1}^N \tilde{a}_l \partial_X A_l(X) , 
\end{equation}
since the derivatives of the basis functions are known analytically.

However, instead of directly 
storing and manipulating the coefficients $\tilde{a}_l$
up to some desired order $N$ in $l$, we make use of the fact
that (for the basis functions of interest) we can derive $\tilde{a}_l$
from the values of $u(X)$ at certain collocation points.
If $u(X)$ is known to have the values $u(X_i) = u_i$ at the 
collocation points $X_i$ for $i=1,2,...,N$ it is possible
to invert the $N$ equations
\begin{equation}
u_i  = \sum_{l=1}^N  \tilde{a}_l A_l(X_i) 
\end{equation}
and to exactly solve for the $N$ coefficients $\tilde{a}_l$ in terms 
of the $u_i$. The location of the different collocation
points depends on the basis functions used. For example, for
Fourier expansions the collocation points have to be equally spaced
in the $X$-interval considered.

This approach of storing the field's values $u_i$ at the collocation
points is called a pseudo-spectral collocation method.
This method has the advantage that non-linear terms such as 
$u(X_i)^2$ can be computed from simple multiplications. Also,
the fields at each point can be evolved forward in time by a
simple method of lines integrator such as Runge-Kutta or iterative
Crank-Nicholson (ICN).

The generalization to 3 dimensions, is straight forward
and can be summarized by
\begin{equation}
u(X_i,Y_j,Z_k) 
= \sum_{l,m,n} \tilde{d}_{lmn} A_l(X_i) B_m(Y_j) C_n(Z_k).
\end{equation}
I.e. we are using basis functions which are products
of functions that depend only on one coordinate. Note that the
three coordinates $X$, $Y$ and $Z$ need not be Cartesian coordinates.

For this, paper we we have chosen $X$, $Y$ and $Z$ to be the spherical
coordinates $r$, $\theta$ and $\phi$. I.e. standard Cartesian
coordinates are given by
\begin{eqnarray}
x &=& r \sin(\theta) \cos(\phi) \\
y &=& r \sin(\theta) \sin(\phi) \\
z &=& r \cos(\theta) .
\end{eqnarray}
As basis functions we will use the Chebyshev polynomials 
\begin{equation}
\label{rbasisfunc}
A_l(r_i) = \cos\left(l 
\arccos\left(\frac{2r_i -R_{out}-R_{in}}{R_{in}-R_{out}} \right)\right)
\end{equation}
in the radial direction, where $R_{in}$ and $R_{out}$ stand for
the inner and outer radii of our numerical domain.
For the angular directions we use the Fourier basis 
\begin{equation}
\label{thetabasisfunc}
B_m(\theta_j) = \frac{1}{N_{\theta}} e^{-i  m \theta_j}
\end{equation}
and
\begin{equation}
\label{phibasisfunc}
C_n(\phi_k) =\frac{1}{N_{\phi}} e^{-i  n \phi_k} .
\end{equation}
The collocation points are chosen to be
\begin{eqnarray}
\label{r_i}
r_i      &=& \frac{R_{in}-R_{out}}{2}\cos\left(\frac{i \pi}{N_{r} - 1}\right)
	    +\frac{R_{in}+R_{out}}{2} \\
\label{theta_j}
\theta_j &=& \pi (2j+1)/N_{\theta} \\
\label{phi_k}
\phi_k   &=& 2\pi k/N_{\phi},
\end{eqnarray}
where
\begin{eqnarray}
i &=& 0,1,...,N_{r} -1, \\
j &=& 0,1,...,N_{\theta} -1 \\
k &=& 0,1,...,N_{\phi} -1 .
\end{eqnarray}
Any function $u(r,\theta,\phi)$ can then be expressed in this basis as 
\begin{equation}
\label{SphericalDF0}
u(r_i,\theta_j,\phi_k)
= \sum_{l=0}^{N_r -1} \sum_{m=0}^{N_{\theta}-1} \sum_{n=0}^{N_{\phi}-1}
\tilde{d}_{lmn} A_l(r_i) B_m(\theta_j) C_n(\phi_k) . 
\end{equation}
However, in our code we never really uses this expansion to
compute all the coefficients $\tilde{d}_{lmn}$. Rather, we only
ever expand in one direction and instead use
\begin{eqnarray}
\label{SphericalDF_r}
u(r_i,\theta_j,\phi_k) 
&=& \sum_{l=0}^{N_r -1} \tilde{a}_{l}(\theta_j, \phi_k) A_l(r_i) , \\
\label{SphericalDF_theta}
u(r_i,\theta_j,\phi_k)
&=& \sum_{m=0}^{N_{\theta}-1} \tilde{b}_{l}(r_i, \phi_k) B_m(\theta_j) , \\
\label{SphericalDF_phi}
u(r_i,\theta_j,\phi_k)
&=& \sum_{n=0}^{N_{\phi}-1} \tilde{c}_{l}(r_i, \theta_j) C_n(\phi_k) ,
\end{eqnarray}
to compute the coefficients $\tilde{a}_{l}(\theta_j, \phi_k)$,
$\tilde{b}_{l}(r_i, \phi_k)$ or $\tilde{c}_{l}(r_i, \theta_j)$
along a line in the $r$-, $\theta$- or $\phi$-direction. 
As we will see, this suffices to compute partial derivatives
in any coordinate direction.

Note that we always choose $N_{\theta}$ to be even, in order to 
assure that collocation points in the $\theta$-direction
(see Eq.~(\ref{theta_j})) do not occur at $\theta=0$ and $\theta=\pi$.
In this way the coordinate singularities at the poles are avoided.
Notice that, since both $\theta$ and $\phi$ are between 0 and $2\pi$, we
have a double covering of the entire domain. This double covering
is due to the fact that we use Fourier expansions in both angles. 

Spherical coordinates also have a coordinate
singularity at $r=0$. In this paper we avoid this
singularity by simply choosing $R_{in}>0$ so that ${r_i}$ in Eq.~(\ref{r_i})
is always larger than zero. This is the appropriate choice
for simulating a single black hole using excision.
From Eq.~(\ref{r_i}) and we see that the collocation points
for $i=0$ and $i=N_r -1$ are located on the inner and outer boundaries
of the numerical domain. Hence it is straightforward to
impose boundary conditions there.

\subsection{Results for non-linear scalar waves}

In order to test our approach we have performed tests with
a non-linear wave equation for a scalar field $\psi$.
When written as a first order in time and second order in space 
system, the evolution equations in Cartesian coordinate are
\begin{eqnarray}
\label{scalarpsi}
\partial_t \psi &=& \Pi \\ 
\label{scalarPi}
\partial_t \Pi &=& (\partial_x^2 + \partial_y^2 + \partial_z^2) \psi 
- \lambda \frac{\psi^3}{1+\psi^2} ,
\end{eqnarray}
where $\lambda$ parametrizes the non-linearity.
This is also how we have implemented this equation in our computer code,
with the caveat that the Cartesian derivatives are computed using
the expansions (\ref{SphericalDF_r}), (\ref{SphericalDF_theta})
and (\ref{SphericalDF_phi}). By this we mean that 
before each timestep derivatives like $\partial_x \psi$ are computed using
\begin{equation}
\partial_x \psi =
 \frac{\partial r}{\partial x} \partial_r \psi
+\frac{\partial \theta}{\partial x} \partial_{\theta} \psi
+\frac{\partial \phi}{\partial x} \partial_{\phi} \psi,
\end{equation}
where $\partial_r \psi$, $\partial_{\theta} \psi$ and $\partial_{\phi} \psi$
can easily be obtained from the expansions in 
Eqs.~(\ref{SphericalDF_r}), (\ref{SphericalDF_theta})
and (\ref{SphericalDF_phi})
since the derivatives of $A_l(r)$, $B_m(\theta)$ and  $C_n(\phi)$
are known analytically.
In order to evolve forward in time we can use any method
of lines integrator. In this paper we will only present results
obtained with a second order in time ICN scheme.
We have, however, also implemented and tested 
several different Runge-Kutta schemes.
In terms of stability all these integrators work equally well
for the cases discussed below, so that the ICN results
suffice for our purposes.

The numerical domain for the scalar wave test is a spherical shell
with an inner radius of $R_{in} = 2$ and an outer radius
of $R_{out} = 52$. Since the grid points are equally spaced
in $\theta$ and $\phi$, the smallest grid spacings occur 
near the poles in the $\phi$-direction. This smallest grid spacing
is of order 
\begin{equation}
h \sim R_{in} \frac{2\pi^2}{ N_{\theta} N_{\phi}} . 
\end{equation}
In order to not violate the Currant condition the
time step $\Delta t$ has to be chosen to be of order $h$ as well.
This means that we have to take very small time steps, so that
the simulations take longer. But it also means that the error
in the ICN time integrator will be only of order
$\Delta t^2 \sim h^2 \sim (N_{\theta} N_{\phi})^{-2}$.

As boundary condition at $r=R_{in}$ we always use
\begin{equation}
u(t,R_{in},\theta,\phi) = u(t=0,R_{in},\theta,\phi) ,
\end{equation}
where $u$ stands for either $\psi$ or $\Pi$.
At $r=R_{out}$ we have tested both the constant boundary condition
\begin{equation}
\label{constBC}
u(t,R_{out},\theta,\phi) = u(t=0,R_{out},\theta,\phi) 
\end{equation}
and also the radiative boundary condition
\begin{eqnarray}
\label{radiativeBC}
\partial_t u(t,R_{out},\theta,\phi) = -v \Big[& &
 \frac{u(t,R_{out},\theta,\phi)}{r} \nonumber \\
&+&\frac{x}{r} \partial_x u(t,R_{out},\theta,\phi) \nonumber \\
&+&\frac{y}{r} \partial_y u(t,R_{out},\theta,\phi) \nonumber \\
&+&\frac{z}{r} \partial_z u(t,R_{out},\theta,\phi) \Big] .
\end{eqnarray}
For our scalar field example we use $v=1$, and apply
this boundary condition to both $\psi$ and $\Pi$ directly,
without decomposing $\psi$ and $\Pi$ into incoming and outgoing
modes at the boundaries.

As initial data for the scalar field we use an inward moving
Gaussian profile given by
\begin{eqnarray}
\psi &=& \frac{A}{r} e^{
-\frac{(x-x_0)^2}{2\sigma_x^2} - \frac{(y-y_0)^2}{2\sigma_y^2} 
-\frac{(z-z_0)^2}{2\sigma_z^2}  },\\
\Pi  &=& -\frac{A}{r} e^{
-\frac{(x-x_0)^2}{2\sigma_x^2} - \frac{(y-y_0)^2}{2\sigma_y^2}
-\frac{(z-z_0)^2}{2\sigma_z^2}  } \nonumber \\
&& \left[ \frac{(x-x_0)x}{\sigma_x^2 r} + \frac{(y-y_0)y}{\sigma_y r}
         +\frac{(z-z_0)z}{\sigma_z^2 r} \right] ,
\end{eqnarray}
where 
\begin{equation}
A=1, \ \ x_0 = 27, \ \ y_0 = z_0 = 0, \ \
\sigma_x = 4, \ \ \sigma_y = \sigma_z = 10
\end{equation}

\begin{figure}
\includegraphics[scale=0.34,clip=true]{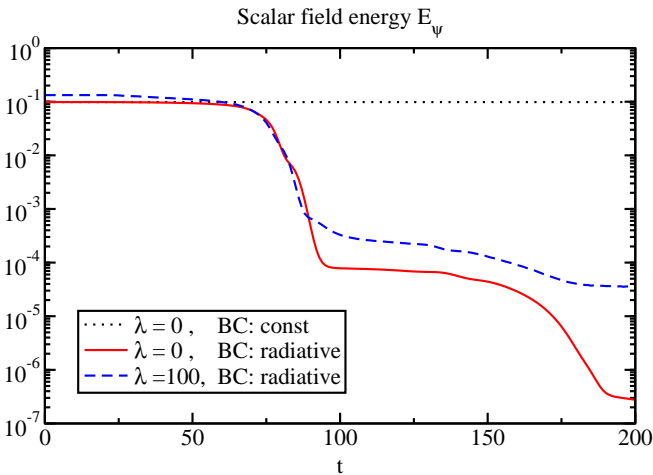} 
\caption{\label{Epsi_vs_t}
The plot shows the scalar field energy inside our numerical domain
for three different cases. 
If the fields are kept constant at the outer boundary 
no energy can leave the domain and thus the energy
remains constant (dotted line). 
If we use the radiative outer boundary condition (\ref{radiativeBC}),
waves can leave the domain and thus the energy
decreases over time. The solid line shows this decrease for the 
linear ($\lambda=0$) case, while the broken line depicts 
the non-linear case ($\lambda=100$).
In all three cases we have used $N_{r}=N_{\theta}=30$ and $N_{\phi}=29$.
}
\end{figure}
Figure \ref{Epsi_vs_t} shows the scalar field energy $E_{\psi}$ 
inside our numerical domain, computed from a volume integral
over the energy density
\begin{eqnarray}
\rho = \frac{\Pi^2}{2} 
&+& \frac{(\partial_x \psi)^2 + (\partial_y \psi)^2 
   + (\partial_z \psi)^2}{2} \nonumber \\
&+& \frac{\lambda}{2} \left[\psi^2 - \log\left(1+\psi^2\right)\right] .
\end{eqnarray}
The plot shows $E_{\psi}$ for different outer boundary conditions
and also for different $\lambda$. The dotted line shows
the result if we use $\lambda=0$, together with the outer boundary
condition (\ref{constBC}). In that case the resulting linear waves
are reflected by the boundaries so that the energy remains constant.
The solid line depicts the $\lambda=0$ case with the radiative
outer boundary condition (\ref{radiativeBC}). Since the initial
wave profile is moving inward, it takes about one crossing time
until the wave starts leaving the grid. After that (around $t\sim 100$)
the energy starts dropping. The same qualitative behavior occurs 
in the non-linear case for $\lambda=100$. 
Note that even though we used identical initial data in all cases, the
energy $E_{\psi}$ in the non-linear case is initially about $35\%$ larger.
Thus for $\lambda=100$ the non-linear terms are not negligible
and are by no means a small perturbation. This means that the 
simple radiative outer boundary condition (\ref{radiativeBC})
is not only capable of propagating non-linear waves off the grid, it also
does not seem to lead to any instabilities in the scalar wave tests
performed by us. This result is interesting since an analogous outer
boundary condition is also used in almost all stable black hole evolution
codes which use the BSSN system together with a finite differencing method.

\begin{figure}
\includegraphics[scale=0.33,clip=true]{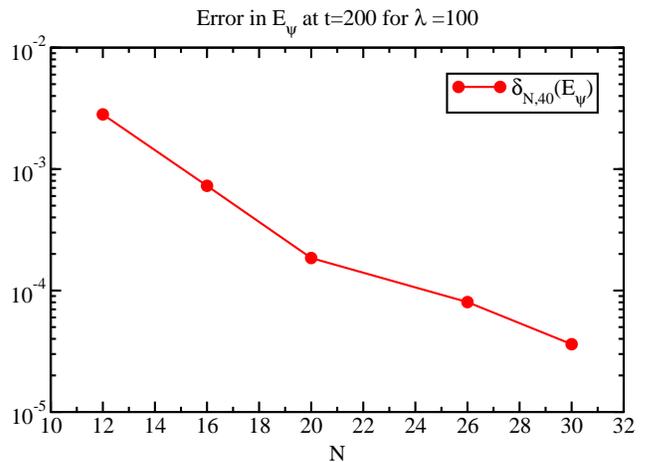} 
\caption{\label{Epsi_t200_conv}
This plot shows the error in $E_{\psi}$ at $t=200$, for the non-linear
wave equation with the radiative outer boundary condition. One can see that
the error decreases exponentially with the number of grid points $N$.
}
\end{figure}

\begin{figure}
\includegraphics[scale=0.33,clip=true]{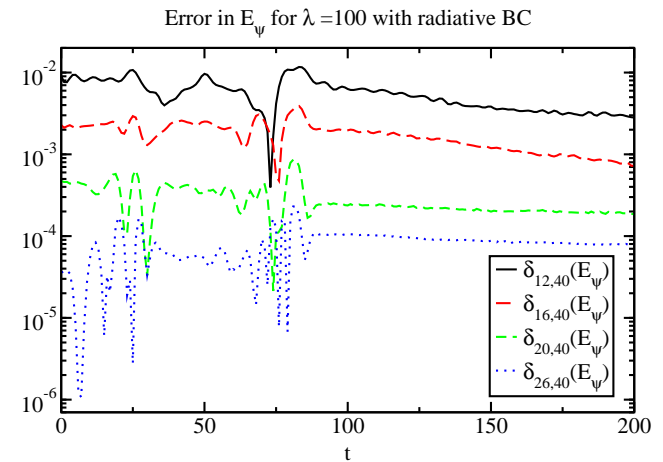}
\caption{\label{Epsi_conv}
This plot shows the error in $E_{\psi}$ versus time, for the the non-linear
wave equation with the radiative outer boundary condition.
}
\end{figure}

\begin{figure}
\includegraphics[scale=0.33,clip=true]{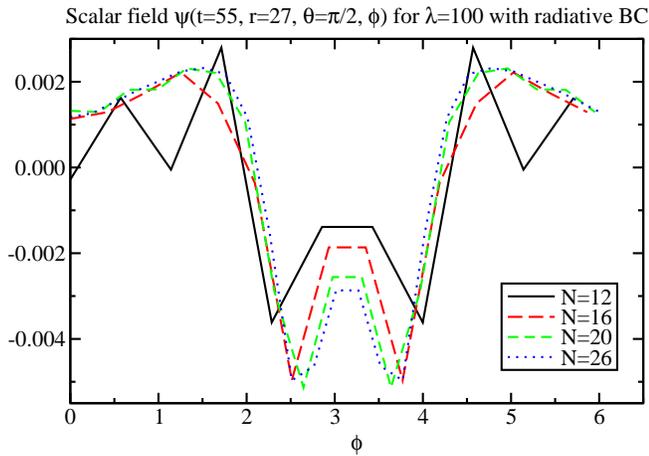}
\caption{\label{psi_conv}
This plot shows the scalar field $\psi$ along an 
equatorial circle ($\theta=\pi/2$) with radius $r=27$.
Shown are snapshots of $\psi$ at $t=55$ for different
numbers of grid points $N_{r}=N_{\theta}=N_{\phi}+1=N$.
}
\end{figure}

In Fig. \ref{Epsi_t200_conv} we demonstrate that our code 
exhibits the expected geometric convergence if we increase
the number of grid points. 
We show the error in $E_{\psi}$ at $t=200$, for the the non-linear
wave equation with radiative outer boundary condition.
Since no analytical solution was available to us, the error in
the energy plotted here is defined by
\begin{equation}
\delta_{N,40}(E_{\psi}) = |E_{\psi,N} -E_{\psi,40}| ,
\end{equation}
where $E_{\psi,N}$ is the $N$th order spectral approximation
to $E_{\psi}$ obtained by using $N_{r}=N_{\theta}=N_{\phi}+1=N$.
In Fig. \ref{Epsi_conv} we show the time evolution of this error.
We can clearly see how the error decreases with an increasing $N$.
Figure \ref{psi_conv} shows the scalar field itself at time $t=55$
along the $\phi$-direction, for $r=27$ and $\theta=\pi/2$.
This time was chosen such that possible noise form the inner and outer
boundaries has had time to reach $r=27$. Also note that during that time
the wave which was initially localized around 
($r=27$, $\theta=\pi/2$, $\phi=0$) has had time to reach
the point ($r=27$, $\theta=\pi/2$, $\phi=\pi$).
We see that for $N=12$ points the wave still looks quite jagged, but
it converges to a smooth result for higher $N$.

\subsection{Analysis of the well-posedness of 
the scalar wave boundary conditions}

As already mentioned the boundary conditions used above are standard, in the
sense that they have been used in almost all black hole simulations with the
BSSN system so far. Like in the scalar field example above, they are applied
to all evolved fields of the BSSN system, without decomposing them into
incoming or outgoing modes at the boundary. Because of this it is not
immediately obvious whether these boundary conditions are mathematically
well-posed. However, at least in the scalar field case well-posedness can be
demonstrated easily. 

Scalar wave equations of the form (\ref{scalarpsi})
and (\ref{scalarPi}) are well studied. For the sake of completeness
we will here quote some results of Gundlach and
Martin-Garcia~\cite{Gundlach:2004ri} who have studied the
well-posedness of systems that are first order in time and
second order in space. First note that the non-linearity
in Eq.~(\ref{scalarPi}) does not affect well-posedness.
Hence our Eqs.~(\ref{scalarpsi}) and (\ref{scalarPi})
are completely analogous to Eqs.~(1) and (2) in \cite{Gundlach:2004ri}.
According to~\cite{Gundlach:2004ri}, the 
second order characteristic variables for direction $n_i$ are given by
\begin{eqnarray}
U_{\pm} &=& \Pi \pm \partial_n \psi \\
U_{A} &=& \partial_A \psi,
\end{eqnarray}
where $\partial_n$ is the derivative in direction $n_i$ and $\partial_A$
are the derivatives normal to $n_i$.
Here $U_{+}$ and $U_{-}$ are the incoming and outgoing modes,
while the $U_{A}$ are zero speed modes.
By using energy estimates, Gundlach and
Martin-Garcia~\cite{Gundlach:2004ri} show that the problem is well-posed
if boundary conditions of the form
\begin{eqnarray}
\label{GundlachBC}
U_{+} = \kappa U_{-} + f
\end{eqnarray}
are imposed on the incoming mode only, where $|\kappa| \leq 1$
and $f$ is a given function.

We will now show that the boundary conditions (\ref{constBC}) and
(\ref{radiativeBC}) lead to a condition of the form (\ref{GundlachBC}).

Let us start with condition (\ref{constBC}) and assume that our
initial data satisfy
$\Pi(t=0,R_{out},\theta,\phi) = \psi(t=0,R_{out},\theta,\phi) = 0 $.
Then Eq.~(\ref{constBC}) leads to $\Pi(t,R_{out},\theta,\phi) = 0$ 
and $\psi(t,R_{out},\theta,\phi) = 0$.
Now $\Pi(t,R_{out},\theta,\phi) = 0$ is completely equivalent
to Eq.~(\ref{GundlachBC}) for $\kappa=-1$ and $f=0$.
So if we had only this one boundary condition, the problem would certainly
be well posed. In this case we should compute $\psi$ at the boundary
by using the evolution equation (\ref{scalarpsi}). However, if we do so,
we immediately obtain that $\psi(t,R_{out},\theta,\phi) = 0$, since 
$\Pi(t,R_{out},\theta,\phi) = 0$.
Thus imposing $\psi(t,R_{out},\theta,\phi) = 0$ is redundant but
completely consistent with the evolution equations.
Hence our boundary condition (\ref{constBC}) is equivalent to
the well posed condition (\ref{GundlachBC}) with $\kappa=-1$ and $f=0$.
Note that in our actual numerical simulations we have really
set $\Pi(t=0,R_{out},\theta,\phi) = \psi(t=0,R_{out},\theta,\phi) = 0$.
I.e. at the boundary we do not really have a Gaussian profile.
However, since our Gaussian is numerically almost indistinguishable from
zero at the boundary anyway, any discontinuities introduced by this
procedure are well below our numerical accuracy.

Next consider the radiative condition (\ref{radiativeBC}). 
For $\psi$ it implies
$\partial_t \psi  = - [ \partial_n \psi + \psi/R_{out}] $.
If we use only this condition for $\psi$ together with the
evolution equation (\ref{scalarpsi}), we can replace
$\partial_t \psi$ by $\Pi$. 
In this case we can compute $\psi$ from the evolution
equation (\ref{scalarpsi}) and our only boundary condition is
\begin{equation}
\label{PiCond}
\Pi  = - [ \partial_n \psi + \psi/R_{out}] .
\end{equation}
This boundary condition, however, is again of the 
form (\ref{GundlachBC}) with $\kappa=0$ and $f=-\psi/R_{out}$.
Thus in this case the problem would be well posed.

Now, the second condition coming from Eq.~(\ref{radiativeBC}), i.e. 
\begin{equation}
\label{rad_Pi}
\partial_t \Pi  = - [ \partial_n \Pi + \Pi/R_{out}] 
\end{equation}
is already implied by the above procedure of using only
the boundary condition (\ref{PiCond}) together with the evolution
equation (\ref{scalarpsi}). (To arrive at Eq.~(\ref{rad_Pi})
we can simply take the time derivative of Eq.~(\ref{PiCond}) and 
use Eq.~(\ref{scalarpsi}) to replace time derivatives of $\psi$.)
Hence the radiative conditions (\ref{radiativeBC}) are simply
an alternate but equivalent way of setting the fields at the
boundary. Since this alternate way can be derived from the well-posed
boundary condition (\ref{PiCond}), it is also well-posed.

In summary, the constant as well as the radiative 
boundary conditions of Eqs.~(\ref{constBC}) and (\ref{radiativeBC})
are both equivalent to imposing conditions only on the 
the incoming scalar wave modes. Thus the scalar wave system 
is well-posed with either one of the two boundary conditions.

\section{Evolving a single black hole with the BSSN system}
\label{sgrid_BSSN}

After having demonstrated that our methods lead to stable
and convergent results for scalar waves we will now turn to evolving
a single black hole example using the BSSN system~\cite{Baumgarte:1998te}.

In the case of BSSN, the $3$-metric $g_{ij}$ is written as
\begin{equation}
g_{ij} = e^{4\phi} \tilde{\gamma}_{ij}
\end{equation}
where the conformal metric $\tilde{\gamma}_{ij}$ has unit determinant.
In addition, the extra variable
\begin{equation}
\tilde{\Gamma}^i 
= \tilde{\gamma}^{ij} \tilde{\gamma}^{kl} \tilde{\gamma}_{jk,l}
\end{equation}
is introduced where $\tilde{\gamma}^{ij}$ is the inverse of
the conformal metric.
Furthermore, the extrinsic curvature is split into its trace 
free part $\tilde{A}_{ij}$ and its trace $K$, and given by
\begin{equation}
K_{ij} = e^{4\phi} 
         \left( \tilde{A}_{ij} + \frac{K}{3} \tilde{\gamma}_{ij} \right) .
\end{equation}
In terms of these variables the evolution equation are
\begin{eqnarray}
\label{dgdt}
\partial_t \tilde{\gamma}_{ij} 
&=& - 2 \alpha \tilde{A}_{ij} + \pounds_{\beta}\tilde{\gamma}_{ij} \\
\partial_t \phi 
&=& \frac{1}{6} \left( -\alpha K + D_i \beta^i \right) \\
\label{dGdt}
\partial_t \tilde{\Gamma}^i
&=& 
    - 2 \alpha \left(  \frac{2}{3} \tilde{\gamma}^{ij} D_j K 
              - 6 \tilde{A}^{ij} D_j \phi
              - \tilde{\Gamma}^{i}_{jk} \tilde{A}^{jk} \right) \nonumber \\
& & -2 \tilde{A}^{ij} D_j \alpha 
    -\xi ( \tilde{\Gamma}^i
       -\tilde{\gamma}^{jk} \tilde{\Gamma}^{i}_{jk}) \beta^l_{,l} 
    + \tilde{\gamma}^{jk} \beta^{i}_{,jk} v\nonumber \\
& & + \frac{1}{3} \tilde{\gamma}^{ij} \beta^k_{,kj}
    - \tilde{\Gamma}^{j} \beta^i_{,j} 
    + \frac{2}{3} \tilde{\Gamma}^{i} \beta^k_{,k} 
    + \beta^j \tilde{\Gamma}^{i}_{,j} \\
\label{dAdt}
\partial_t \tilde{A}_{ij}
&=& e^{-4\phi} 
    \left[ - D_i D_j \alpha 
           + \alpha \left(  \tilde{R}_{ij} 
                          + R^{\phi}_{ij}\right) \right]^{TF} \nonumber \\
& & + \alpha (K \tilde{A}_{ij} - 2 \tilde{A}_{ik}\tilde{A}^{k}_{j}) 
    + \pounds_{\beta}\tilde{A}_{ij} \\
\label{dKdt}
\partial_t K
&=& - D^i D_i \alpha 
    + \alpha \left( \tilde{A}^{ij}\tilde{A}_{ij}+ \frac{1}{3} K^2 \right) 
    + \pounds_{\beta}K .
\end{eqnarray}
Here the superscript $TF$ in Eq.~(\ref{dAdt})
denotes the trace free part and $D_i$ is the derivative 
operator compatible with the $3$-metric $g_{ij}$. Notice that
$\tilde{\Gamma}^{i}_{jk}$ and $\tilde{R}_{ij}$ are the 
Christoffel symbol and Ricci tensor associated with
the conformal metric $\tilde{\gamma}_{ij}$, while
\begin{eqnarray}
R^{\phi}_{ij}
&=& -2 \tilde{D}_i \tilde{D}_j \phi 
    -2 \tilde{\gamma}_{ij} \tilde{D}^l \tilde{D}_l \phi \nonumber \\
& & +4 (\tilde{D}_i \phi) (\tilde{D}_j \phi)
    -4 \tilde{\gamma}_{ij} (\tilde{D}^l \phi) (\tilde{D}_l \phi),
\end{eqnarray}
where $\tilde{D}_i$ is the derivative operator compatible with the 
conformal metric $\tilde{\gamma}_{ij}$.
For all the results reported here we have set the constant $\xi$ in
Eq.~(\ref{dGdt}) to $\xi = 4/3$.


We should also note that in order to ensure that $\tilde{A}_{ij}$
remains traceless during our numerical evolution, we subtract
any trace due to numerical errors after each evolution step
from $\tilde{A}_{ij}$.

As initial data we use the metric and extrinsic curvature of 
a Schwarzschild black hole of mass $M$ in Kerr-Schild coordinates.
Thus initially we have
\begin{eqnarray}
g_{ij} &=& \delta_{ij} + 2 H l_i l_j \\
K_{ij} &=& \alpha [ l_{i} H_{,j} + l_{j} H_{,i} + H l_{i,j} + H l_{j,i} \nonumber \\
       & &+ 2 H^2 (l_{i} l_{k} l_{j,k} + l_{j} l_{k} l_{i,k}) 
          + 2 H l_{i} l_{j} l_{k} H_{,k} ] ,
\end{eqnarray}
where
\begin{eqnarray}
H = M/r \nonumber \\
l^i = x^i/r . 
\end{eqnarray}
The indices of $l^i$ can be lowered with the flat metric.

As initial lapse and shift we use
\begin{eqnarray}
\label{init_lapse}
\alpha = 1/\sqrt{1 + 2 H}, \\
\label{init_shift}
\beta^i = 2 l^i H/(1 + 2 H) .
\end{eqnarray}
If this lapse and shift are kept constant during the evolution
the right hand sides of Eqs.~(\ref{dgdt})-(\ref{dKdt}) are all zero
and thus all the BSSN variables are constant.
All the results presented in this paper are indeed obtained with
a constant shift equal to what is given in Eq.~(\ref{init_shift}).
The lapse, however, is treated somewhat differently and we have
tried two approaches. 
Firstly, following~\cite{Arbona99,Alcubierre00a,Yo02a,Sperhake:2003fc} 
we have experimented with a so called $1+\log$-lapse given by
\begin{equation}
\label{lapse_evo}
\partial_t \alpha = D_i \beta^i - \alpha K .
\end{equation}
The second approach is to introduce a densitized lapse given by
\begin{equation}
\label{lapseDensity}
q = \frac{\alpha}{\sqrt{g}} ,
\end{equation}
where $g$ is the determinant of the 3-metric.
As in~\cite{Sperhake:2003fc} this densitized lapse $q$ is then
evolved instead of $\alpha$, which is simply computed using 
Eq.~(\ref{lapseDensity}).
In our particular case we impose
\begin{equation}
\label{constDensLapse}
\partial_t q = 0 ,
\end{equation}
and thus we keep the densitized lapse constant during the evolution.
Analytically both Eq.~(\ref{lapse_evo}) and Eq.~(\ref{constDensLapse})
lead to a constant lapse for the Kerr-Schild initial data used here.
However, as we will see below, numerically both cases differ.

The BSSN system of evolution equations (Eqs.~(\ref{dgdt})-(\ref{dKdt}))
is first order in time and second order in space, just as the scalar
wave example considered before. We will thus use the same numerical
methods. As our numerical domain we use a spherical shell, which
extends from $R_{in}=1.85M$ to $R_{out}=16M$, with collocation
points given by Eqs.~(\ref{r_i})-(\ref{phi_k}). 
On this shell we will use the Chebyshev basis (\ref{rbasisfunc})
in the radial direction and the Fourier basis (\ref{thetabasisfunc})
and (\ref{phibasisfunc}) in the angular directions.
As before, we will evolve the Cartesian components of all variables
on this spherical grid without rewriting the evolution equations
in spherical coordinates. 

The horizon of the black hole is at $r=2M$, so that the inner boundary
at $r=R_{in}=1.85M$ is inside the horizon, which justifies 
excising the region $r<R_{in}$. We treat this inner 
boundary like a pure outflow boundary and do not impose any 
boundary condition there. Apart from using a spectral method
this is the one point where we really have to do something
that differs from what is usually done when the BSSN system is
evolved using finite differencing methods. Almost all
finite differencing BSSN simulations performed so far
either avoid excision altogether or, if excision is used, apply the 
so called ``simple excision'' boundary condition~\cite{Alcubierre00a},
where the time derivatives of variables at the inner boundary are
obtained by copying the time derivatives from neighboring grid points.
Even though one might worry about spoiling geometric convergence,
this ``simple excision'' algorithm can in principle also be applied
within our spectral method using collocation points. 
The results, however, are disastrous
and the simulation crashes after only a few $M$ of evolution time.
In the continuum limit the ``simple excision'' technique
corresponds to a Von Neumann boundary condition, which can
also be applied within our spectral approach. This inner 
boundary condition, however, also leads to exponential blow up
after only a few $M$ of evolution time. Hence the simplest
prescription which can be used without decomposing our variables
into incoming and outgoing modes, 
is to impose no boundary condition at $r=R_{in}$.
Fortunately this option is easily implemented within our spectral
approach, and amounts to treating the points at $r=R_{in}$
in the same way as interior points by simply using the BSSN 
evolution equations at these points as well.

At the outer boundary at $r=R_{out}=16M$
we have experimented with both constant and radiative boundary conditions.
In the constant case we simply apply Eq.~(\ref{constBC})
to all evolved variables. Analytically this is the correct
boundary condition since with our choices for lapse and shift
nothing should evolve, so that all fields indeed remain constant
at the boundary. 
We have also tested radiative boundary conditions. 
Following~\cite{Yo02a} we have applied a modified radiative
boundary condition to our evolved variables. The modification
consists of applying condition~(\ref{radiativeBC}) 
to the difference between the evolved variables and the analytic solution
given by the initial data. As in~\cite{Yo02a} we 
have applied this modified radiative boundary condition to all
variables except $\tilde{\Gamma}^{i}$, which 
was kept constant at the outer boundary.

In the following we will present results for three different
combinations of the boundary conditions and lapse choices described above.
\begin{figure}
\includegraphics[scale=0.33,clip=true]{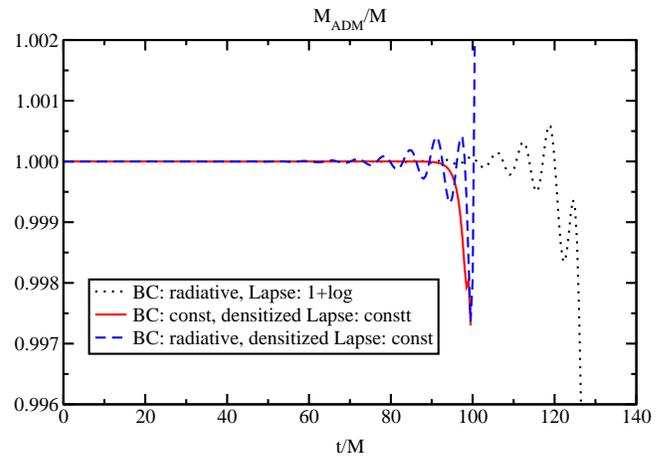} 
\caption{\label{E_ADM}
ADM mass versus time for different choices of lapse and outer boundary
conditions. The short dashed line is for the radiative outer boundary
condition together with a $1+\log$ lapse. This simulation crashes at
$t=137M$. The solid line depicts the result when all variables are kept
constant at the outer boundary and a constant densitized lapse is used. In
this case the run dies at $t=99M$. The long dashed line shows the case for
the radiative boundary condition together with a constant densitized lapse.
Here a crash occurs at $t=100M$. All results shown in this figure were
obtained using $N_{r}=30$, $N_{\theta}=14$ and $N_{\phi}=13$.
}
\end{figure}
In Fig. \ref{E_ADM} we show the ADM mass (see e.g. \cite{Wald84}) versus time.
In principle the ADM mass should stay constant for all time. However,
due to exponentially growing instabilities all our runs eventually 
crash and the ADM mass becomes more and more inaccurate over time.
The two dashed lines are obtained when we use the radiative outer boundary
conditions described above. The short dashed line is for the $1+\log$
lapse of Eq.~(\ref{lapse_evo}). This simulation lasts the longest.
The long dashed line corresponds to a simulation where the densitized
lapse of Eq.~(\ref{lapseDensity}) is kept constant.
The solid line is for a constant densitized lapse together with
constant outer boundary conditions. This simulation crashes at about the
same time as the corresponding one with radiative boundary 
conditions (long dashed line).

Let us discuss next how these results depend on the resolution
used. To this end it is instructive to study the normalized Hamiltonian
constraint at different resolutions.
\begin{figure}
\includegraphics[scale=0.33,clip=true]{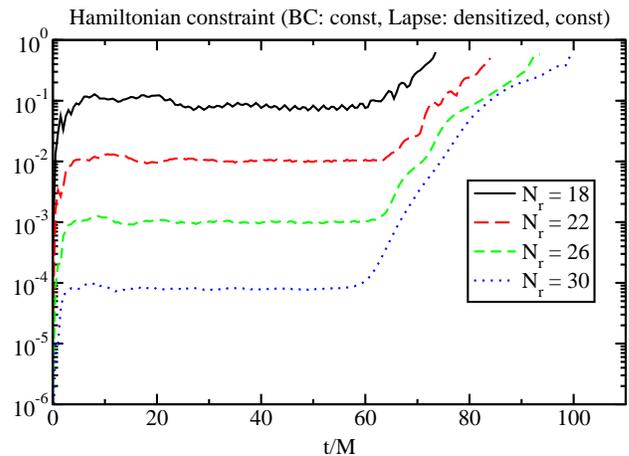} 
\caption{\label{ConstBC_Dens1Const}
The plot shows the $L^2$-norm over the entire grid of the
normalized Hamiltonian constraint versus time for the case of
constant outer boundary conditions and a constant densitized lapse.
In this plot $N_{\theta}-1=N_{\phi}=13$. For increasing $N_r$
we observe geometric convergence in the Hamiltonian until $t \sim 60M$.
}
\end{figure}
Figure \ref{ConstBC_Dens1Const} shows the Hamiltonian for an 
increasing number of radial grid points $N_r$ in the case of
constant outer boundary conditions and a constant densitized lapse.
After an initial increase in the Hamiltonian for the first few $M$,
the curves become flat until instabilities take over at
around $t \sim 60M$. These instabilities eventually lead to a crash.
We obtain geometric convergence until about $t \sim 60M$.
Note that the simulations last longer when the 
resolution is increased.

\begin{figure}
\includegraphics[scale=0.33,clip=true]{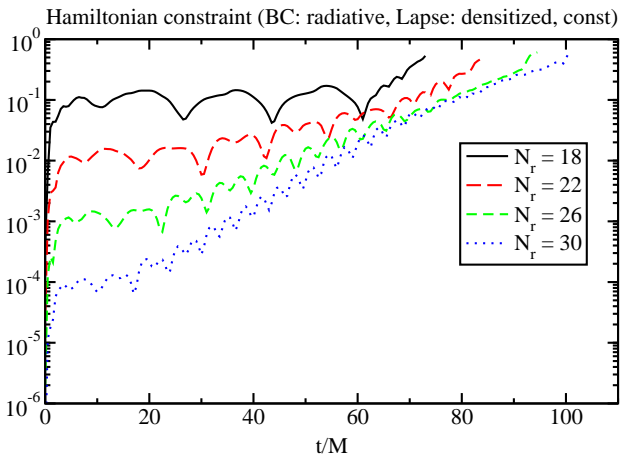} 
\caption{\label{radana_Dens1Const}
The $L^2$-norm of the normalized Hamiltonian constraint versus time 
for the case of radiative outer boundary conditions and a constant 
densitized lapse. For increasing $N_r$ the Hamiltonian converges
until about $t \sim 60M$.
The angular resolution is $N_{\theta}-1=N_{\phi}=13$.
}
\end{figure}

In Fig. \ref{radana_Dens1Const} we show the corresponding results
for the case with radiative outer boundary conditions and a
constant densitized lapse. 
Again the result converges with resolution until about $t \sim 60M$
and simulations run longer for higher resolutions. 
However, this time a flat plateau forms only for a short time 
after the initial increase of Hamiltonian. For example for the highest
resolution the curve is almost flat between $2M$ and $20M$.
After that the Hamiltonian increases exponentially.
This increase is the result of constraint violations
entering through the outer boundary, since the radiative boundary
conditions used here are not constraint preserving.

\begin{figure}
\includegraphics[scale=0.33,clip=true]{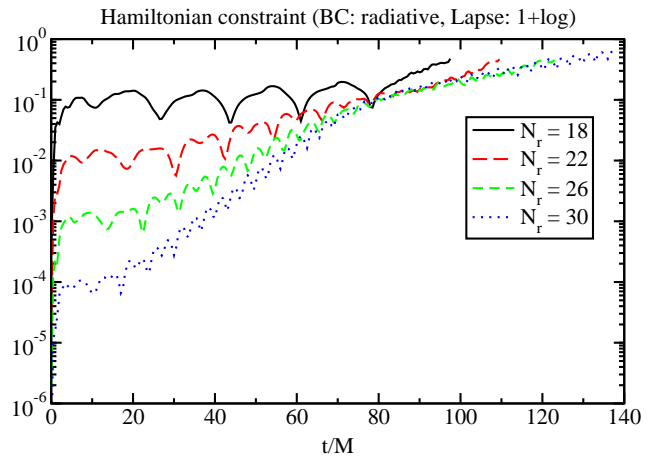} 
\caption{\label{radana}
The $L^2$-norm of the normalized Hamiltonian constraint versus time 
for the case of radiative outer boundary conditions and a $1+\log$
lapse. For increasing $N_r$ the Hamiltonian converges
until about $t \sim 60M$.
The angular resolution is $N_{\theta}-1=N_{\phi}=13$.
}
\end{figure}

Figure \ref{radana} shows our results for the case of $1+\log$ lapse 
together with radiative outer boundary conditions.
The qualitative behavior in this case is the same as for the constant
densitized lapse with radiative outer boundary conditions presented before
in Fig. \ref{radana_Dens1Const}. Note, however, that the time axis
has been extended since these runs last longer.

All plots were obtained for an angular resolution of $N_{\theta}=14$
and $N_{\phi}=13$. All we have done so far was to vary the radial
resolution. Note that the Schwarzschild black hole evolved here is
spherically symmetric. Thus, when the Cartesian components of our evolved
variables are expanded in a Fourier series in $\theta$ or $\phi$, only the
three lowest frequencies contribute, so that only the first three complex
coefficients can be non-zero. Since all our variables are real, the lowest
coefficient must also be real. This implies that the first three complex
coefficients can be described by five real numbers, which corresponds to
five collocation points. Hence if we assume spherical symmetry throughout
the evolution, a resolution of $N_{\theta}=N_{\phi}=5$ would be entirely
sufficient. Recall however, that in order to avoid collocation points at the
poles we require $N_{\theta}$ to be even, and thus the minimum we need is
$N_{\theta}-1=N_{\phi}=5$. If we run our simulations with this minimum
angular resolution non-spherical modes are suppressed and our simulations run
up to 50\% longer than the ones shown above. Since our ultimate goal is to
be able to also evolve non-spherically symmetric situations we have chosen
not to do this. When we increase $N_{\theta}-1=N_{\phi}$ from 5 to a number
between 6 and 10 we find that run-time decreases. Yet for values of
$N_{\theta} \ge 14$ and $N_{\phi} \ge 13$ all results are largely
independent of angular resolution. For this reason
we have presented only results for $N_{\theta}-1=N_{\phi}=13$.

Recall that in the case of finite differencing BSSN simulations with
excision, instabilities were observed when a single black hole is evolved on
a full grid~\cite{Alcubierre00a}. These instabilities disappear when
symmetries such as octant symmetry are imposed. In the present work
no symmetries are imposed. This might be one reason for the observed
instabilities. In fact, here we do not only evolve a full grid, but the
full grid is also covered twice due to our use of Fourier expansions
in both angles. This double covering corresponds to less
symmetries than in usual full grid simulations. Thus one might worry
that it could even lead to additional instabilities, above and
beyond the full grid instability~\cite{Alcubierre00a}.
This is however not the case. 

The double covering can be removed
by using the fact that all evolved variables $u$ have to satisfy
\begin{equation}
\label{resetDC}
u(r,\theta,\phi) = u(r,2\pi-\theta,\phi+\pi) = u(r,2\pi-\theta,\phi-\pi)
\end{equation}
if $\pi \leq \theta < 2\pi$.
Thus if we use an even number of collocation points in both angles
we can simply overwrite all variables at collocation points for which
$\pi \leq \theta<2\pi$ with their values at the corresponding points 
with $0\leq \theta < \pi$, after each time step.
Alternatively, one can also take the average of $u$ at corresponding
points and and replace $u$ at both points with this average.
We have tried both and methods to remove the double covering and
both give the same result. 

\begin{figure}
\includegraphics[scale=0.33,clip=true]{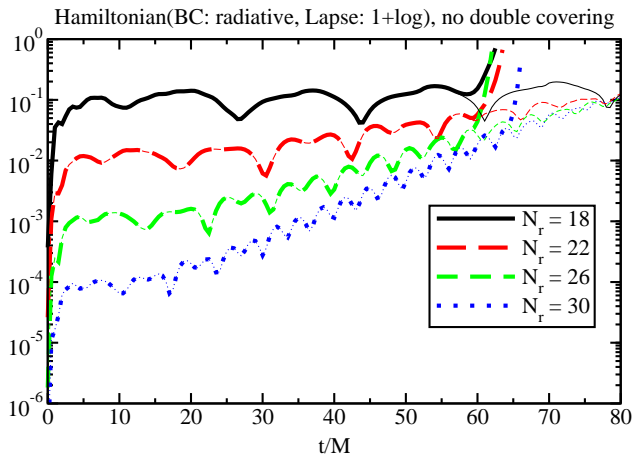}
\caption{\label{radana_resetDC}
The $L^2$-norm of the normalized Hamiltonian constraint versus time
for radiative outer boundary conditions and a $1+\log$
lapse. The thick lines show how the Hamiltonian evolves when we
remove the double covering (using Eq. (\ref{resetDC})). For comparison
the results with double covering (from Fig. \ref{radana})
are shown as thin lines. When the double covering 
is removed the runs crash soon after convergence is lost 
(around $t \sim 60M$), while the runs with double covering
carry on for a while with large errors and without converging.
Note that in the convergent regime there is no noticeable
difference between double covering and single covering, and both
work equally well. The angular resolution is $N_{\theta}=N_{\phi}=14$.
}
\end{figure}
The thick lines in Fig. \ref{radana_resetDC} show the time evolution of the
Hamiltonian constraint for simulations where the the double covering has
been removed by using the averaging procedure described above, after each
time step. For comparison the thin lines also depict the corresponding
results with double covering. As one can see both single and double covering
produce indistinguishable results until convergence is lost at around $60M$.
After that the simulations with single covering blow up quickly, while the
simulations with double covering continue for some time (see Fig.
\ref{radana}). However, after about $60M$ the simulations with double
covering have large errors and do not converge any longer. Hence 
the effective runtime during which we get useful results is equal
for both single and double covering. Thus we conclude that both
single and double covering work equally well and that the double covering
is not the reason for the observed instabilities in the BSSN
evolutions described here.

Another way to remove the double covering would be to expand our variables
in spherical harmonics instead of double Fourier expansions.
This is, however, beyond the scope of this paper.

When working with spectral methods it is quite common to use filters to
suppress instabilities. For example in the case of Fourier expansions one
often used filter consists of zeroing out the top most third of all spectral
coefficients. This so called $2/3$ rule can also be used for Chebyshev
expansions. We would like to point out that no filters were used in the work
presented above. We have, however, tested the simple $2/3$ rule in our case
and found that it has no effect on the observed instabilities.
The only effect we observe is a reduced accuracy when the filters are on.
This reduction is explained by the lower effective resolution when
$1/3$ of all coefficients is zeroed out.

\section{Discussion}
\label{discussion}

We have, for the first time, evolved the BSSN system by means of a
pseudo-spectral collocation method. This method uses a grid of collocation
points that are distributed over a spherical shell. On this grid we evolve
the Cartesian components of all variables. We use Fourier expansions in both
angles and a Chebyshev expansion in the radial direction.
This double Fourier expansion leads to a double covering of the
spherical shell in the sense that there are always two grid points
which correspond to the same physical location in the shell.
The fact that there are two corresponding points where all
evolved variables must have the same value can also be
used to undo the effects of the double covering after each
evolution step. However, as we have seen in Sec.~\ref{sgrid_BSSN},
the simulations with single and double covering yield basically
the same results in the convergent regime.

The BSSN system is second order in space, and we have implemented it as is,
without any first order reductions. The main purpose of this work was to
test this system together with boundary conditions and gauges that have so
far only been used together with finite differencing methods. In particular,
we use black hole excision and either radiative or constant outer boundary
conditions. As gauges we have tested both $1+\log$ lapse and constant
densitized lapse. Depending on these choices we can evolve a single black
hole in Kerr-Schild coordinates for about $100M$. After that the simulation
crashes due to unstable exponentially growing modes.
While this evolution time is not too impressive by today's
standards, it demonstrates that most methods used in codes based
on finite differencing can in principle also by used with spectral methods. 
In fact, our code tests with a non-linear wave equation show
that both constant as well as radiative boundary conditions
can be simply applied to all evolved fields without
decomposing them into ingoing and outgoing modes.
In this scalar wave case all our simulations are stable
and no blowups were observed.
Therefore it seems likely that the observed instabilities
which occur with the BSSN system when we evolve a single black hole,
are not caused by the simple radiative or constant outer boundary
conditions used here. Rather the blowup may due to the use of black
hole excision on a full grid without imposing any symmetries 
such as octant or bitant symmetry. 
Such a full grid instability has also been observed in
finite differencing implementations of BSSN~\cite{Alcubierre00a}
when excision is used.
Indeed, we have not checked whether the version of the BSSN used here has
some incoming modes (e.g. gauge modes) at the excision boundary. If such
modes do exist, they require boundary conditions at the excision surface
and our choice of not imposing any boundary conditions there may be the
cause of the observed instabilities. We leave the study of such modes to
future work.

We would also like to point out that the runtime of about $100M$ in the
simulations described above, was achieved using the standard BSSN system
without any parameter fine tuning. For example, it is well known that one
can modify this system through the addition of constraints on the right hand
side of the evolution equations. These additional terms can be multiplied by
parameters which can be adjusted to extend the runtime.
Recall that when the Einstein-Christoffel system~\cite{Anderson99}
was first implemented~\cite{Kidder01a} 
it ran for only about $40M$ before it crashed.
By adding constraints with parameters this system was extended
into what is now known as the KST system~\cite{Kidder01a}.
After parameter fine tuning this KST system was able to evolve 
single black holes for the much longer time of about 
$8000M$~\cite{Scheel2002a}.
It might be possible to extend the runtime of the BSSN simulations described
in this work in an analogous way. Furthermore, the runtime and accuracy of
our simulations should improve, if the simple constant or radiative boundary
conditions used here are replaced by constraint preserving boundary
conditions.

\begin{acknowledgments}

It is a pleasure to thank Bernd Br\"ugmann, Marcus Ansorg, Mark Scheel 
and Lawrence Kidder for useful discussions about spectral methods.
This work was supported by NSF grant PHY-0555644.
We also acknowledge partial support by the
National Computational Science Alliance under
Grants PHY050012T, PHY050015N, and PHY050016N.

\end{acknowledgments}




\bibliography{references}

\end{document}